\documentclass[conference, letterpaper]{IEEEtran}
\usepackage[letterpaper, left=.625in, right=.625in, bottom=1in, top=0.75in]{geometry}
\IEEEoverridecommandlockouts

\usepackage{setspace}

\usepackage{cite}

\usepackage{lipsum}
\usepackage{multicol}
\usepackage{wrapfig}
\usepackage{sidecap}
\usepackage{graphicx}
\graphicspath{{images/}}
\usepackage{subcaption}
\usepackage{algpseudocode}
\usepackage{algorithm}
\usepackage{amsmath}
\usepackage[dvipsnames]{xcolor}

\usepackage{algorithm}
\usepackage{listings}
\usepackage{color}

\definecolor{dkgreen}{rgb}{0,0.6,0}
\definecolor{gray}{rgb}{0.5,0.5,0.5}
\definecolor{mauve}{rgb}{0.58,0,0.82}

\newcommand{\refsec}[1]{Section~\ref{#1}}
\newcommand{\reffig}[1]{\mbox{Figure~\ref{#1}}}

\newcommand{\etal}{~et~al.}

\usepackage{amssymb}

\newcommand{\linebreakand}{%
  \end{@IEEEauthorhalign}
  \hfill\mbox{}\par
  \mbox{}\hfill\begin{@IEEEauthorhalign}
}

\def\BibTeX{{\rm B\kern-.05em{\sc i\kern-.025em b}\kern-.08em
    T\kern-.1667em\lower.7ex\hbox{E}\kern-.125emX}}

\begin{document}

\title{Cost-Effective Situational Awareness \\Through IoT COTS Radios}

\author{
\IEEEauthorblockN{Batuhan Mekiker}
\IEEEauthorblockA{
\textit{Montana State University}\\
Bozeman, MT, USA\\
batuhan.mekiker@montana.edu}
\and
\IEEEauthorblockN{Alisha Patel}
\IEEEauthorblockA{
\textit{Beartooth Radio Inc.}\\
Los Angeles, CA, USA\\
alisha@beartooth.com}
\and
\IEEEauthorblockN{Mike P. Wittie}
\IEEEauthorblockA{
\textit{Montana State University}\\
Bozeman, MT, USA \\
mike.wittie@montana.edu}
}

\maketitle
\begin{abstract}
Situational awareness~(SA) through encrypted connectivity in modern warfare is critical to battlefield coordination. 
Existing solutions provide infrastructure-less connectivity and situational awareness, but at a high cost, power consumption, and bulky form factors that limit their utility for tactical operations. 
We propose the Beartooth MKII radios as a cost- and power-effective, handy solution for establishing a situational awareness overlay.
To integrate with military information brokers, such as the Tactical Assault Kit~(TAK), we propose the Beartooth Gateway, which creates an IP network between Beartooth radios, Gateways, and other TAK-capable devices.
We show that leveraging Internet of Things~(IoT) Commercial Off The Shelf~(COTS) components for MKII and the Gateway can enhance the availability of SA and meet the user requirements in the field while ensuring compatibility with existing IP networks connected to TAK infrastructure. 
\end{abstract}


\section{Introduction}
\label{sec:introduction}

As we have seen in recent conflicts, the nature of warfare is evolving. The absence of encrypted communications generates an operational disadvantage for large military forces and enables small, highly mobile units to disrupt their advance~\cite{rep:marine}. 
On the other hand, accurate, up-to-date situational awareness~(SA), whose definition can vary depending on the scenario, is crucial. 
For example, for troops on the battlefield, SA might refer to immediate access to markers of enemy troop locations, while for wildfire firefighters, it might entail rapid delivery of polygon-defined areas on the map to describe wildfire containment. 
Regardless of the specifics, SA allows these small units and their command centers to coordinate efforts. 
Consequently, a low-cost, scalable SA solution that can be tailored to different use cases 
becomes an increasingly important force multiplier that first responders and military commanders need in their arsenal.

Mission-critical or  specifically battlefield SA relies on two integrated technologies:
a connectivity layer that provides raw communication links between agents,
and an information layer that disseminates location-based data and operational directives.
Battlefield radios from Silvus~\cite{site:silvus}, Trellisware~\cite{site:trellis}, \mbox{and Persistent Systems~\cite{site:waverelay}} can establish direct as well as multi-hop connectivity among agents.
These radios also integrate with mobile phones running the Android Tactical Assault Kit~(ATAK)~\cite{paper:atak} to create an SA overlay.
Networks built on these radios, however, have several shortcomings. 
First, these purpose-built radios are comparably more expensive than other radios built with commercially-available components, which makes it difficult to deploy them in large numbers.
Unfortunately, the price per unit for these radios is not publicly available. 
However, discussions with users revealed that comparable network sizes would be 5-10 times more costly than networks built with commercial off-the-shelf~(COTS) radio components.
Second, they are power hungry which increases their weight, limits portability, and increases observability~\cite{site:townsend}.
Finally, third, without an edge Tactical Assault Kit~(TAK)~\cite{site:tak} server in the field, SA data dissemination with SA communication systems that use these radios is geographically limited. 
SA data can only be exchanged within the coverage area of the radios, limiting SA to just one or a few local squads.

One way to reduce SA system costs and form factor is to build it with COTS components.
Research on internet-of-things~(IoT) communications has produced several multihop connectivity solutions~\cite{paper:zwave_lora, paper:sigfox, paper:DASH7, paper:zigbee}.
These low-cost radios provide low-bitrate links, which may nevertheless be suitable for SA applications.
A key advantage of these radios is their low cost and spectral efficiency, which leads to low power requirements and small, lightweight form factors.
Despite their benefits, these low-cost radios must also satisfy certain Quality of Service~(QoS), such as low latency, delivery receipts for SA data, and resilience to network disruptions caused by high mobility, to be considered for SA systems and mission-critical applications. 
However, whether resource-limited IoT radios can meet these demands is unclear.

In this paper, we propose an SA solution based on IoT COTS radios.
Specifically, we design the Beartooth MKII radios based on the XBee radio platform~\cite{site:xbee}.
The XBee radio platform provides more than ten miles of range in urban areas and sixty miles in rural areas with a line-of-sight between transceivers.
XBee also supports DigiMesh, a self-forming, self-healing link layer protocol with built-in frequency hopping and routing among more than a hundred nodes in a Low Power Wide Area Network~(LPWAN)~\cite{paper:digimesh}.
In addition, XBee radios are highly power-efficient and draw less than 3\,mA while in standby mode, and 900\,mA during data transmission at 30\,dBm~(1\,W) transmit power~\cite{site:xbee_specs}. 
Finally, the XBee platform is inexpensive with modules easily accessible commercially.

\begin{figure}
    \centering
    \begin{subfigure}[b]{0.63\columnwidth}
        \includegraphics[width=\textwidth]{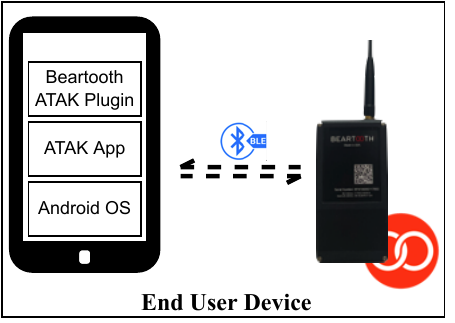}
        \caption{Beartooth ATAK Plugin and MKII radio.}
        \vspace{5pt}
        \label{fig:eud}
    \end{subfigure}
    \begin{subfigure}[b]{0.65\columnwidth}
        \includegraphics[width=\textwidth]{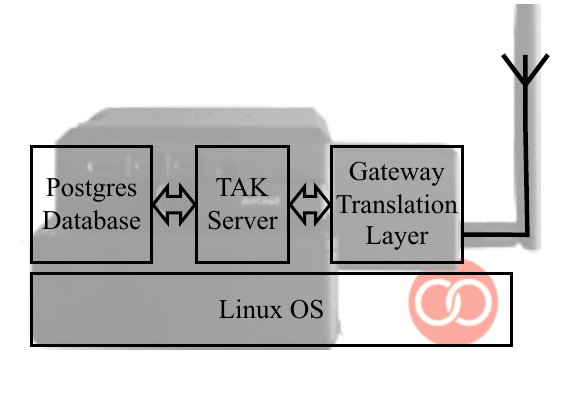}
        \vspace{-20pt}
        \caption{Beartooth Gateway and its software modules.}
        \label{fig:gw}
    \end{subfigure}
    \caption{Bearthooth system design.}
    \vspace{-4mm}
\end{figure}

The Beartooth MKII radio couples an XBee radio with a Bluetooth module that allows it to speak to a mobile phone running ATAK.
To address the bitrate limitation of the XBee platform, we design the Beartooth ATAK Plugin
to rearrange the existing TAK data format known as Cursor-on-Target~(CoT) events into a more compact representation. 
We compress larger data types such as sensor readings, images, and bulk data and we split them for transmission over multiple XBee frames. 
We also use unicast transmissions to reduce time-on-air for larger CoT events, which route frames directly to the destination to reduce the number of data and acknowledgment packets traversing on the network with respect to broadcast frames.
While Beartooth MKII radios exchange our compact frames, the original CoT events still get recreated at the receiver then published to ATAK, allowing us to share SA in a resource-constrained COTS network.
The resulting MKII form factor, shown in \reffig{fig:eud}, is 4.09 x 1.21 x 0.7 inches and weighs in at 6\,oz, including a battery.
The Beartooth MKII radios are as much as 54\% lighter than existing tactical radios discussed in \refsec{sec:related_work}.

\begin{figure*}
    \centering    \includegraphics[width=0.7\textwidth]{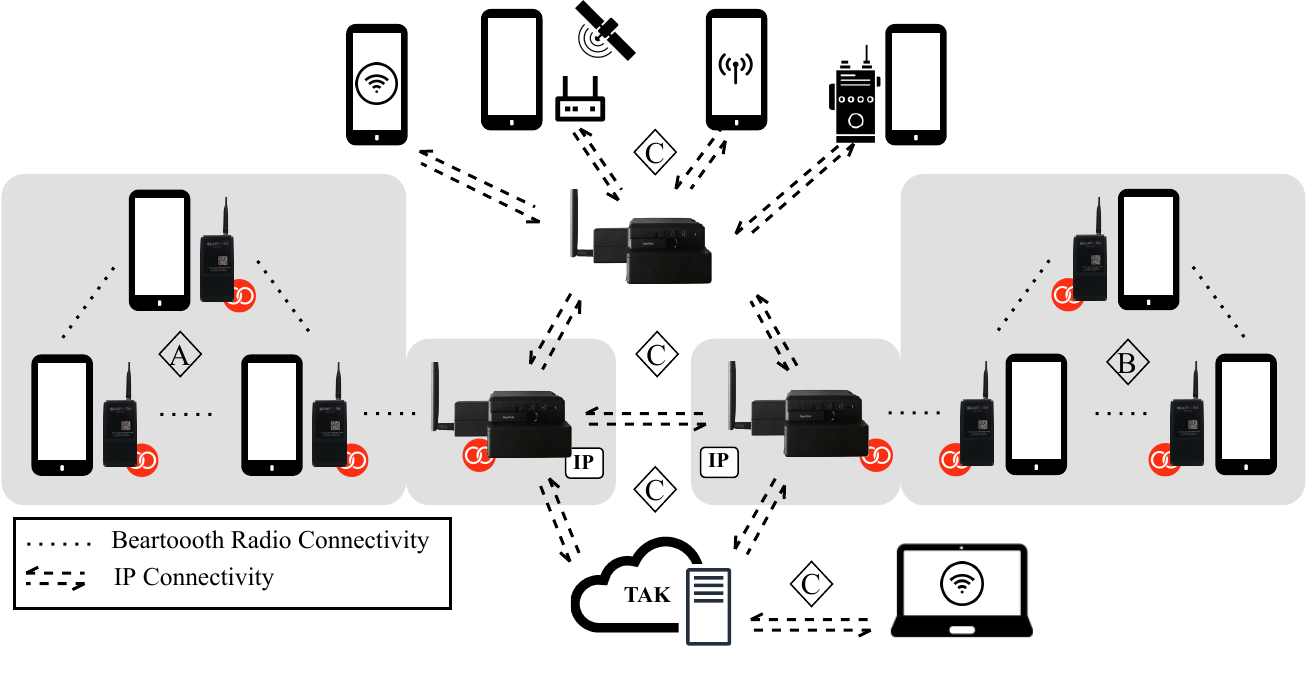}
    \caption{An example deployment of Beartooth network with MKII radios and Gateways, connecting many users spanning across WiFi, cellular~(5G/LTE), satellite connectivity and battlefield radios. } 
    \label{fig:topology}
\end{figure*}

Further, we provide shared SA between squads and commanders through the Beartooth Gateway.
The Gateway is a handheld server as seen on \reffig{fig:gw} that consist of a MKII radio and a translation layer to couple an XBee network to an IP network over cellular, satellite, or other radio technologies, such as the radios from Silvus, Trellisware, Persistent Systems, or Harris. 
IP connectivity allows the sharing of encrypted SA data among multiple squads as well as command centers running a TAK server~\cite{site:tak}.

\reffig{fig:topology} illustrates a possible deployment scenario interconnecting multiple network technologies and operational areas into a shared TAK SA overlay.
In \reffig{fig:topology} we show two Beartooth networks, \textit{A} and \textit{B}, bridged together using two Gateways.
Both Gateways are connected through an IP network such as the internet with the help of a Virtual Private Network~(VPN) denoted as network~\textit{C}. 
An additional Gateway within network \textit{C} serves the TAK information layer to multiple IP-connected devices through various network interfaces such as WiFi, cellular~(5G/LTE), and satellite connectivity.
All three TAK servers within Gateways are federated, allowing Beartooth SA data to flow into the IP network and SA data generated at network \textit{C} to flow into Beartooth Networks \textit{A} and \textit{B}.
A fourth TAK server sits in the cloud computing architecture, is also part of the network \textit{C}, and is federated to Gateway \textit{A} and~\textit{B}. 
Therefore both Beartooth network members and members of IP network can exchange SA and battlefield intelligence. 
Furthermore, commanding officers can observe the whole operation and even send orders through the TAK server in the cloud using a WinTAK~\cite{paper:wintak} device, creating 
communication links that are encrypted between forces in the field and an operation command center.

This paper's main contribution is enabling SA over IoT COTS radios that are compatible with traditional SA systems through the following supporting technical contributions. 

\begin{enumerate}
\item A secure, bandwidth-efficient transport layer for TAK messages over XBee platform
\item A routing layer for TAK messages among Beartooth Gateways and TAK servers
\end{enumerate}


A Beartooth network composed of MKII radios and Gateways delivers a capable SA solution to coordinate squads and command centers over large areas of operation.
The low cost and small form factor of the Beartooth radios make it possible to deploy SA at a large scale and is a powerful force multiplier.


\section{Related Work}
\label{sec:related_work}
To illustrate the need for cheaper, smaller, and more flexibly deployable radios, we discuss the limitations of existing solutions to create connectivity and situational awareness. 
We consider commercially available tactical radios as well as COTS technologies from the IoT space.

In the commercial space, there are three predominant solutions Streamcaster, TSM Shadow, and MPU5 radios from Silvus Technologies, TrellisWare Technologies, and Persistent Systems respectively.
All were developed to provide tactical connectivity and SA for first responders and military forces.

Streamcaster product line from Silvus Technologies uses a proprietary radio module capable of self-forming and self-healing a mesh IP network~\cite{site:silvus}. 
Streamcaster Lite radio 
has a maximum of 1\,W transmit power and 20\,Mbps data rate.
The radio requires 4.8\,W to 17\,W power while transmitting at 1\,W~(30\,dBm), which is a few times more than the Beartooth MKII radio. 
The Streamcaster Lite weighs in at 10.4\,oz and takes up more than twice the physical space of a Beartooth MKII.

Similar to Streamcaster, TrellisWare's TSM Shadow handheld radio encloses a proprietary radio module~\cite{site:trellis}. 
It has a transmit power of up to 2\,W with a maximum data rate of 16\,Mbps.
Although there is no publicly available information about power consumption, it would be safe to assume TSM Shadow would consume similar power as Streamcaster Lite. 
It weighs around 11.3\,oz, an ounce more than Streamcaster Lite however, volume-wise, it sits between the Beartooth MKII and the Streamcaster Lite.

MPU5, compared to Streamcaster Lite and TSM Shadow, is a more complete and capable radio system with a full-fledged mobile CPU and Android Operating System on board~\cite{site:waverelay}. 
Without the battery module, It weighs around 13\,oz.
MPU5 has a wide variety of radio modules for many frequency bands with data rates up to 150\,Mbps. 
With varying radio modules for different frequency bands, its transmit power varies between 4\,W to 10\,W, and its power consumption varies between 30\,W and 50\,W.

All three radios have IP network capability and provide sufficient bandwidth to support hundreds of devices sharing real-time CoT events.
However, their physical footprint and cost are significantly higher than Beartooth MKII radios.
While all three radios do support IP networks and can handle TAK-formatted SA data, thus ensuring compatibility with the TAK ecosystem out-of-the-box, their functionality is geographically limited without an edge TAK server in the field. 
They can exchange local SA data within their coverage area, but to transmit SA data beyond this local network, an edge TAK server is required. 
Given this requirement for an edge server to extend coverage, it could be more beneficial to opt for physically smaller, easily concealable, cheaper, and therefore more expendable radios, such as Beartooth radios and Gateways.
We are interested in bringing cost-effective COTS devices to SA systems by adopting radio modules developed for a wide variety of IoT applications.
The IoT solution space commonly utilizes solutions like Z-Wave~\cite{paper:zwave_lora}, SigFox~\cite{paper:sigfox}, \mbox{DASH-7~\cite{paper:DASH7},} LoRa~\cite{paper:zwave_lora}, and ZigBee~\cite{paper:zigbee}, which provide comparatively low-cost and power-efficient options.
Z-Wave and SigFox can support
infrequent data transmission at a low data rate over long distances. 
However, due to their inability to meet the QoS requirements of SA and TAK traffic requires such as low latency and resilience to high mobility, these solutions are disqualified from use in mobile SA systems.
\mbox{DASH-7} and LoRa can sustain communication links with longer distances and sufficient maximum raw data rate of 200\,Kbps and 37.5\,Kbps respectively for single-hop networks while keeping power consumption low. 
However, no official or third-party link layer protocol implements a multi-hop self-forming, self-healing mesh network with performance SA systems require. 
For LoRa, in earlier work, we found building a mesh network challenging due to limited bandwidth, concluding that a second radio module is needed to extend coverage beyond two hops~\cite{paper:BM_lora}.
Finally, ZigBee's use in both sub-GHz and 2.4\,GHz bands and compatibility with other ZigBee devices enhance its IoT use case. 
However, its network deployment is cumbersome, because it requires extensive planning of different radio roles.

Similar to ZigBee, DigiMesh, available on XBee platform, supports sub-GHz and 2.4\,GHz ISM bands, with similar performance metrics like bandwidth, data rates, latency, and power consumption as discussed by Khalifeh\etal~\cite{paper:zigbee}.
However, DigiMesh's key advantage is its simplified network role structure and its self-forming, self-healing mesh network which allows the network to be fluid and highly mobile. 
Unlike ZigBee, which requires distinct coordinator, router, and end-device roles, thereby adding complexity to deployment and maintenance, DigiMesh operates with a singular network role, easing deployment, maintenance, and network extension.

While DigiMesh and other IoT solutions hold certain advantages, they are not without drawbacks. 
The most notable limitations of DigiMesh are its inability to support data rates at Mbps levels and the lack of support for IP network connectivity. 
Despite these shortcomings, we show that core SA data types—such as text, geolocation, markers, polygonal shapes, and voice—can be effectively supported with proper on air data management. 
Furthermore, by using gateways, we can translate Beartooth SA data into traditional TAK formatted IP SA data, ensuring compatibility with existing infrastructure. Taking into account both the benefits and drawbacks, we have chosen the XBee platform and DigiMesh as our link layer protocol over other IoT COTS solutions.


\section{Beartooth Network}
\label{sec:bt_gateways}

To understand the effectiveness of the Beartooth network, we discuss the software components that both MKII and Gateway require for effective communication. 
We then elaborate on the types of situational awareness data that the MKII radios and Gateway can serve to the Beartooth Network and the IP-layer network. 
\vspace{-1mm}
\subsection{Network Elements}
\label{sec:netw_elem}
The MKIIs are highly mobile and power-efficient handheld radios that forces in the field use to communicate within a local region. 
Gateways within Beartooth networks are edge servers bringing TAK capability to the field and integrating troops using MKII radios to a larger TAK SA overlay.

\subsubsection{MKII}
To limit the form factor of MKIIs we design them with limited computational power and place much of the transport layer logic on the connected phone. 
There are a few reasons we opt to use this design decision. 
First, designing and building a radio without a powerful processing unit is cheaper and less complex.
Second, having a power-hungry processing unit affects the power consumption and thus overall battery life of MKII radios.
Last but not least, since the mobile phone does all the required processing, the protocol design is quite flexible and easily upgradable.
If and when we decide to change how we encode data into packets, we can do so by modifying the phone plugin without the necessity to update the MKII radio firmware.

One of the key design decision involves efficiently serializing and transporting data in a resource-limited network. 
We use Google's Protocol Buffers~(ProtoBuf) library~\cite{site:protobuf}, allowing platform-neutral serialization of any data structure. 
ProtoBuf uses schemas, simplifying the encoding of various data fields and types.

We define all Beartooth SA packet within a ProtoBuf schema where we defined fields such as {\small \texttt{sourceUid}}, 
{\small \texttt{destinationUid}}, 
{\small \texttt{messageType}}, 
{\small \texttt{textPayload}}, 
{\small \texttt{messageUid}} and 
{\small \texttt{locPayload}}.
While designing the Beartooth packet schema, we focused on the limited network resources. 
CoT data format, being XML-based and designed for IP networks, is not optimally suited for resource-limited networks due to its inherent verbosity. 
Hence, we extract only the necessary data from the CoT event and form it into the Beartooth packet format, thereby minimizing its footprint.
Once we have the proper packet format, ProtoBuf encodes necessary data into a series of bytes, then passes those bytes to the Beartooth radio interface.



The Beartooth MKII plugin for ATAK is a lightweight platform that connects to Beartooth MKII radio using Bluetooth Low Energy (BLE) to send various SA data such as text, location, markers, shapes, pictures, and voice messages.
Together with the MKII radio, we design the plugin to be a robust and secure communications platform. 
As the default option, it auto-shares the user location with the entire network and a team leader to monitor the squad members. 
So that the team leader can take appropriate actions. 
In addition to the types mentioned above of SA messages, the plugin is also capable of sending casualty evacuation~(casevac), navigation routes for walking, flying, and driving, icons for a wide variety of first responder missions, elevation, range and bearing.
Once the plugin is connected to MKII through the BLE, the plugin configures the radio in the background to user preferences, where the user can set network channel and encryption settings.

Contrary to other radios, MKII radios coupled with the Beartooth ATAK plugin do not directly transmit CoT events due to limited network resources. 
When the user triggers a SA event, such as a marker on the map in ATAK, the plugin puts necessary data to a Beartooth SA frame using the ProtoBuf schema, serializes it, and sends it to the BLE connection. 
Then, BLE queues the data for the radio in MKII to be transmitted.
End-user devices receiving the Beartooth SA frame pass the data to the plugin. 
Using the ATAK API, the plugin only then recreates CoT events at the receiver device to be shown on the ATAK user interface. 

We also implemented a network scanning tool enabling end-users to plan and monitor their Beartooth network deployment. 
The tool uses MKII radios to ping all devices within a local DigiMesh network, visualizing RSSI values, distances, hops, and overall network health on the Beartooth ATAK plugin. 
This helps users optimize radio deployment for stronger communication links and manage the Beartooth network with ease.

\subsubsection{Gateway}
The Gateway, shown in \reffig{fig:gw}, consists of three individual modules: Virtual Private Network~(VPN), TAK Server, and Gateway Translation Layer.
We use ZeroTier VPN~\cite{site:zero} which is lightweight, easy to set up peer-to-peer VPN solution that provides static IP addresses for TAK servers within the Gateways and IP-connected end user devices. 
Furthermore, we use TAK server version 4.8 
optimized for Single Board Computers~(SBC).
It allows us to bring the TAK server to the edge of the network and connect multiple local networks through the TAK's federation protocol. 
While the federation protocol's details fall outside this paper's scope, it is important to note that it facilitates the secure exchange of all, or selected, SA data between authenticated TAK servers through encrypted TLS links. 
If there is any connection interruption, it is handled within the TAK.

\begin{figure*}[]
    \centering
    \includegraphics[width=0.7\textwidth]{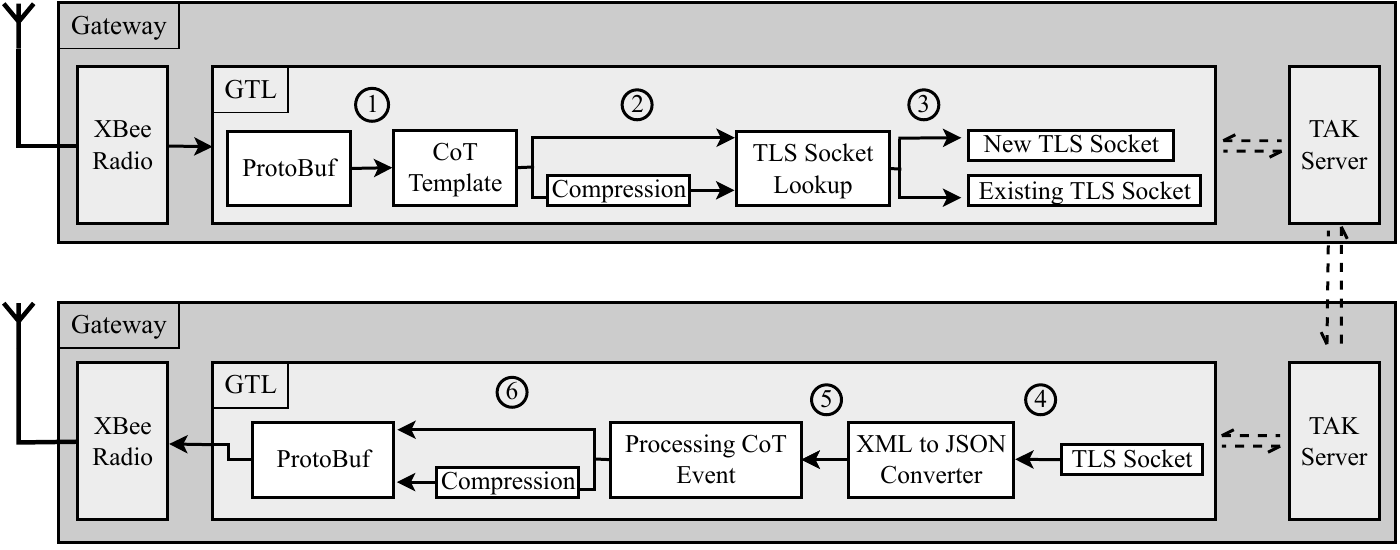}
    \caption{Data progression within Gateway and GTL showing various steps and software modules. }
    \label{fig:dataflow}
\end{figure*}

\vspace{-0.5mm}
\subsection{Gateway Translation Layer}


We design Gateway Translation Layer~(GTL) which is the novel solution that links Beartooth MKII devices to the TAK Server and the IP network through an XBee radio connected to its serial port. 
GTL enables bi-directional SA data communication, as illustrated in \reffig{fig:dataflow}. 
The following sections describe GTL's processing of incoming and outgoing data for both directions in detail.

\subsubsection{Data Flow: Beartooth to IP-network}
We show the data flow from Beartooth MKII devices to an IP network through a Gateway in \reffig{fig:dataflow}.
The GTL, using the pre-determined Beartooth's ProtoBuf schema, parses the incoming SA data frame from the Beartooth network and determines the data and CoT event type. 
Depending on the type, the GTL process SA data within the DigiMesh frame as shown in \textit{step 1} and forms a valid CoT event out of one of the pre-configured CoT event formats representing the original SA data frames.
The pre-configured CoT events are templates that help us generate valid CoT events by pluging in relevant Beartooth SA data.

After forming a valid CoT event, depending on the CoT event type the GTL can route fully formed CoT event to the compression engine in \textit{step 2} to reduce message time-on-air. 
Only medium-sized CoT events such as marker, routes and shapes and CoT events carrying bulk data such as pictures and compressed data packets go through the compression engine. 
In \textit{step 3} the GTL looks up for active TLS sockets based on sender address.
If there is an established TLS socket, the GTL recycles that active TLS socket to submit CoT event to TAK. 
Otherwise, the GTL creates a TLS socket from available client certificates generated during the build process and stored at the Gateway.
Having multiple certificates helps us keep latency low and TLS connections sticky for ongoing transmissions and recently active users.
Once GTL establishes a TLS connection with the TAK server, it submits the generated CoT event and stores the connection for a brief period.
Depending on the type and destination of the event, TAK Server triggers a set of procedures determining how CoT events get processed and routed.

\subsubsection{Data Flow: IP to Beartooth Network}
Conversely, as seen on bottom section of \reffig{fig:dataflow}, when the data path is from the IP network to Beartooth Network, SA data is either routed from a TAK Server as seen in~\reffig{fig:dataflow} or it is received by the bottom TAK server via a device such as an ATAK that is connected to the this TAK server through a TLS connection.   
Then TAK relays SA data received in CoT format to GTL.
In \textit{step 4} the GTL converts XML-based CoT event into a JSON object for ease of accessibility. 
GTL then processes the JSON CoT events by categorizing them by the event type on \textit{step 5}.
As we discussed, if the CoT event type is what we classify as medium-sized or bulk data, the SA data goes through a compression procedure.
Similarly, GTL forms the Beartooth SA data frame from either compressed or uncompressed SA data using Beartooth's ProtoBuf schema in \textit{step 6}. 
Then the GTL passes the serialized data in bytes onto to the radio module using the serial port. 
The radio module then transmits data either by unicasting or broadcasting depending on the data type.

\vspace{-0.5mm}
\subsection{Supported Data Types}
The Gateway supports two-way communication between the Beartooth network and IP network. 
Therefore the Gateway support data flows from IP-network to the Beartooth network, the Beartooth network to IP-network, or the Beartooth network to the Beartooth network with Gateways in between.
The Gateway also supports various CoT event types and meets the required QoS for the SA through the GTL.
The next couple of paragraphs shows SA data types Beartooth network and the Gateway supports.

\textbf{\textit{Small-sized packets (up to 256B)}} are three distinct data types: text, geo-location, and acknowledgment. 
Text and acknowledgment messages are CoT events used for communication and delivery receipt confirmation within the TAK overlay. 
Geo-location messages, another type of CoT event, contain latitude, longitude, and altitude data, enabling end-users to share their location.
    
\textbf{\textit{Medium-sized packets (256B to a few KB)}} are CoT events that are map overlays and can be represented similarly such as markers, routes, and shapes. 
Due to DigiMesh's 256B frame payload limit, we transmit larger CoT events in a series of frames. 
A compression engine deflates XML-based CoT events to minimize spectrum usage and frame count. 
The GTL stores received frames, compiling complete CoT events once all frames are received. 
With built-in retransmission protocol, if a frame is lost during transmission, GTL can request retransmission through a selective acknowledgment mechanism to ensure reliability and accurate CoT event compilation.

\textbf{\textit{Bulk Data Packets (up to 25KB)}} are {\small \texttt{.zip}} files that include an XML-based manifest, guiding TAK on the enclosed data type and processing requirements. 
The Beartooth network supports various data formats, such as images, map overlays, icon sets, and TAK server configurations, given the data does not exceed 25KB. 
The transmission strategy mirrors that of medium-sized packets, using a series of frames for data delivery including the packet loss and retransmission protocol. 






\section{Evaluation}
\label{sec:eval}

In this section, we evaluate the application layer performance of Beartooth Network as a whole including GTL.
Beartooth devices utilize the DigiMesh protocol from the XBee platform. 
As a result, they inherit its performance.
A detailed performance evaluation of DigiMesh is beyond the scope of this paper~(see work by Khalifeh\etal~\cite{paper:zigbee} for an overview). 

\textbf{\textit{Latency}} \hspace{1.5mm} 
To demonstrate Beartooth MKII radios and Gateways' capabilities in the SA use case, we evaluated their performance on supported data types by taking end-to-end message latency measurements. 
These measurements encapsulate the entire SA data life-cycle, from dissemination in the Beartooth Network to processing by the GTL and TAK server. 
Our setup included one MKII radio, a phone~(Samsung Galaxy XCover) with the Beartooth ATAK plugin, an IP-connected phone~(Samsung Galaxy XCover) with ATAK, and a Gateway. 
The latency measurements, reflecting single-hop latency for each data type, were based on timestamps from transmission start till SA data received by the IP-connected phone, with each experiment repeated a hundred times.

First, we evaluate small-sized packets such as text and location data. 
We observe the latency
is 0.73\,s on average with variance around 0.02\,s. 
For medium-sized packets where CoT event types such as markers, routes, and shapes we have two scenarios. 
In the first scenario, markers and simple polygonal shapes~(e.g. 2-3 vertices), the latency is 2.15\,s on average.
In the second scenario markers and shapes have increased detail and complexity~(e.g. \textit{casevac} marker with all the attributes or polygonal shape with 4+ vertices), the latency reaches upwards of 2.5\,s. 
Finally, for bulk data, we had a compressed file enclosing an image sized 6\,KB. 
We observe that the latency is around 20.1\,s on average.
These numbers met the user QoS requirements of field trial evaluations.


\textbf{\textit{Scalability}}
\hspace{1.5mm} We evaluated the application layer scalability of our Beartooth network in several military exercises, where it underwent significant stress, demonstrated by a large number of active devices and SA data packet transmissions. In one instance, we used around 100 Beartooth end-user devices, each updating their location every five seconds and exchanging SA text messages. Both field forces and authorities evaluating our network solution reported no issues with reliability or performance throughout the exercise, and we observed no service interruptions. 
These results highlight the Beartooth network's scalability and its ability to maintain functionality and dependability under high-demand conditions.

\section{Conclusion and Future Work}
\label{sec:conc_and_fw}

In this paper we presented a COTS network deployment with Beartooth MKII radios and Gateways to provide SA to first responders and military forces spanning multiple Beartooth and IP networks.
We believe extending the usage of COTS radios in the form of Beartooth MKII radios and introducing Gateway to interconnect IP networks through TAK can provide scalable SA at low-cost, with a compact form factor.
Our experiments show Beartooth devices can support real-time text, location as well as more complex SA data types such as markers, shapes and bulk data with tolerable latency and QoS.
Users have provided overwhelmingly positive feedback, with satisfaction regarding the currently supported network size and SA data types.
With several active war zone deployments and military training, the MKII radios and Gateways are currently being battle tested. 

Our future plans include integrating the Beartooth Plugin with various systems and ATAK plugins, increasing compatibility by integrating diverse sensor applications. 
We aim to enhance Gateway performance via multi-threading and parallel processing. 
Developing a web interface is already underway for simplified Gateway management and WebTAK use, expanding SA capabilities to all web browser-capable platforms.


\bibliographystyle{./references/IEEEtran}
\bibliography{./references/gateway}

\begin{thebibliography}{10}
\providecommand{\url}[1]{#1}
\csname url@samestyle\endcsname
\providecommand{\newblock}{\relax}
\providecommand{\bibinfo}[2]{#2}
\providecommand{\BIBentrySTDinterwordspacing}{\spaceskip=0pt\relax}
\providecommand{\BIBentryALTinterwordstretchfactor}{4}
\providecommand{\BIBentryALTinterwordspacing}{\spaceskip=\fontdimen2\font plus
\BIBentryALTinterwordstretchfactor\fontdimen3\font minus
  \fontdimen4\font\relax}
\providecommand{\BIBforeignlanguage}[2]{{%
\expandafter\ifx\csname l@#1\endcsname\relax
\typeout{** WARNING: IEEEtran.bst: No hyphenation pattern has been}%
\typeout{** loaded for the language `#1'. Using the pattern for}%
\typeout{** the default language instead.}%
\else
\language=\csname l@#1\endcsname
\fi
#2}}
\providecommand{\BIBdecl}{\relax}
\BIBdecl

\bibitem{rep:marine}
\BIBentryALTinterwordspacing
``{Strategy for the Long Haul: The {US Marine Corps} Fleet Marine Forces for
  the 21st Century},'' Report, CSBA Online, 2008. [Online]. Available:
  \url{https://tinyurl.com/CSBA-Report}
\BIBentrySTDinterwordspacing

\bibitem{site:silvus}
``{STREAMCASTER RADIOS },'' Spec. Sheet, Silvus Technologies, 2021,
  https://tinyurl.com/silvus-scaster-specs.

\bibitem{site:trellis}
\BIBentryALTinterwordspacing
``{{TSM} Shadow Product Datasheet},'' Spec. Sheet, TrellisWare Technologies,
  2021. [Online]. Available: \url{https://tinyurl.com/trellis-tsm-shadow}
\BIBentrySTDinterwordspacing

\bibitem{site:waverelay}
\BIBentryALTinterwordspacing
``{MPU5 Spec. Sheet},'' Spec. Sheet, Persistent Systems, 2021. [Online].
  Available: \url{https://tinyurl.com/mpu5-spec-sheet}
\BIBentrySTDinterwordspacing

\bibitem{paper:atak}
K.~Usbeck, M.~Gillen, J.~Loyall, A.~Gronosky, J.~Sterling, R.~Kohler,
  K.~Hanlon, A.~Scally, R.~Newkirk, and D.~Canestrare, ``{Improving situation
  awareness with the {Android Team Awareness Kit~(ATAK)}},'' in \emph{(C3I)
  Technologies for Homeland Security, Defense, and Law Enforcement XIV}.\hskip
  1em plus 0.5em minus 0.4em\relax SPIE, 2015.

\bibitem{site:townsend}
\BIBentryALTinterwordspacing
``{Lighter, faster, smaller equipment needed for soldiers to win, says {Gen.
  Townsend}},'' News Article, US Army, 2018. [Online]. Available:
  \url{https://tinyurl.com/army-Gen-Townsend}
\BIBentrySTDinterwordspacing

\bibitem{site:tak}
\BIBentryALTinterwordspacing
``{Team Awareness Kit: Tactical Situational Awareness Solution},'' Fact Sheet,
  DHS, 2020. [Online]. Available: \url{https://tinyurl.com/tak-fact-sheet}
\BIBentrySTDinterwordspacing

\bibitem{paper:zwave_lora}
S.~Al-Sarawi, M.~Anbar \emph{et~al.}, ``{Internet of Things (IoT) communication
  protocols: Review},'' in \emph{{IEEE} Information Technology (ICIT)}, May
  2017.

\bibitem{paper:sigfox}
\BIBentryALTinterwordspacing
``{Sigfox Technology Overview},'' Jul. 2019. [Online]. Available:
  \url{https://tinyurl.com/what-sigfox}
\BIBentrySTDinterwordspacing

\bibitem{paper:DASH7}
P.~Mach, Z.~Becvar, and T.~Vanek, ``{Internet of Mobile Things: Overview of
  LoRaWAN, DASH7, and NB-IoT in LPWANs Standards and Supported Mobility},''
  \emph{{IEEE} Commun. Surveys Tuts.}, vol.~21, no.~2, Oct. 2018.

\bibitem{paper:zigbee}
A.~Khalifeh, H.~Salah, S.~Alouneh, A.~Al-Assaf, and K.~Darabkh, ``Performance
  evaluation of {DigiMesh} and {ZigBee} wireless mesh networks,'' in
  \emph{Wireless Communications, Signal Processing and Networking}, 2018.

\bibitem{site:xbee}
\BIBentryALTinterwordspacing
``{Digi XBee Ecosystem},'' Online Resource, DIGI. [Online]. Available:
  \url{https://tinyurl.com/digi-xbee}
\BIBentrySTDinterwordspacing

\bibitem{paper:digimesh}
\BIBentryALTinterwordspacing
``{Wireless Mesh Networking: ZigBee vs. DigiMesh},'' White Paper, DIGI, 2018.
  [Online]. Available: \url{https://tinyurl.com/zigbee-vs-digimesh}
\BIBentrySTDinterwordspacing

\bibitem{site:xbee_specs}
\BIBentryALTinterwordspacing
``{Digi XBee SX 900 RF Module},'' 2022. [Online]. Available:
  \url{https://tinyurl.com/xbee-specs}
\BIBentrySTDinterwordspacing

\bibitem{paper:wintak}
\BIBentryALTinterwordspacing
``{{Team Awareness Kit (TAK)} - Enhancing Homeland Security Enterprise
  Collaboration on the Mobile Edge},'' White Paper, DHS and S\&T, 2019.
  [Online]. Available: \url{https://tinyurl.com/tak-dhs}
\BIBentrySTDinterwordspacing

\bibitem{paper:BM_lora}
B.~Mekiker, M.~P. Wittie, J.~Jones, and M.~Monaghan, ``Beartooth relay
  protocol: Supporting real-time application streams with dynamically allocated
  data reservations over {LoRa},'' in \emph{Computer Communications and
  Networks (ICCCN)}, 2021.

\bibitem{site:protobuf}
\BIBentryALTinterwordspacing
``{Protocol Buffers - Google's data interchange format},'' GitHub Repo, Google.
  [Online]. Available: \url{https://tinyurl.com/protobuf-github}
\BIBentrySTDinterwordspacing

\bibitem{site:zero}
\BIBentryALTinterwordspacing
``{Protocol Design Whitepaper},'' White Paper, ZeroTier, 2018. [Online].
  Available: \url{https://tinyurl.com/zerotier-protocol}
\BIBentrySTDinterwordspacing

\end{thebibliography}
\end{document}